\documentclass[11pt]{article}
\usepackage{a4}
\usepackage{psfig}
\usepackage{epsfig}
\usepackage{epsf}
\usepackage{rotating}
\usepackage{citesort}       
\usepackage{amssymb}
\usepackage{graphicx}
\def\lsim{\raise0.3ex\hbox{$\;<$\kern-0.75em\raise-1.1ex\hbox{$\sim\;$}}}
\def\gsim{\raise0.3ex\hbox{$\;>$\kern-0.75em\raise-1.1ex\hbox{$\sim\;$}}}

\newcommand{\vecc}[1]{\mbox{\boldmath $#1$}}
\newcommand{\vR}{\vecc{R}}            

\newcommand{\matr}[1]{\mbox{#1}}
\newcommand{\eV}{\matr{eV} }

\def\CCSNO{\mbox{\rm CC-SNO} }
\def\ESSNO{\mbox{\rm ES-SNO} }
\def\NCSNO{\mbox{\rm NC-SNO} }
\newcommand{\boron}{\mbox{${}^8$B\ }}
\newcommand{\fluxunit}{\mbox{$10^6$ cm$^{-2}$ s$^{-1}$ }}

\def\npb#1#2#3{    { Nucl. Phys. }{\bf B #1} (19#2) #3}
\def\npbps#1#2#3{  { Nucl. Phys. }(Proc. Suppl.){\bf B #1} (19#2) #3}
\def\plb#1#2#3{    { Phys. Lett. }{\bf B #1} (19#2) #3}
\def\prd#1#2#3{    { Phys. Rev. }{\bf D #1} (19#2) #3}

\def\prl#1#2#3{    { Phys. Rev. Lett. }{\bf #1} (19#2) #3}

\begin{document}
\begin{flushright}
{\small
CERN-TH-2002-170\\
IFUM-716/FT\\
FTUAM-02-378
}
\end{flushright}

\begin{center}
{ \Large \bf Determination of Neutrino mixing parameters after 
 SNO oscillation evidence}\\[0.2cm]

{\large P.~Aliani$^{a\star}$, V.~Antonelli$^{a\star}$,
 R.~Ferrari$^{a\star}$, M.~Picariello$^{a\star}$,
E.~Torrente-Lujan$^{abc\star}$
\\[2mm]
$^a$ {\small\sl Dip. di Fisica, Univ. di Milano} and
{\small\sl  INFN Sezione di Milano, \\ Via Celoria 16, Milano, Italy}\\
$^b$ {\small\sl Dept. Fisica Teorica C-XI, 
Univ. Autonoma de Madrid, 28049 Madrid, Spain,}\\
$^c$ {\small\sl CERN TH-Division, CH-1202 Geneve}\\
}

\end{center}

\begin{abstract}
An updated analysis of all available neutrino oscillation evidence 
 in Solar experiments (SK day and night spectra, global rates from 
 Homestake, SAGE and GALLEX) 
 including the latest $SNO  CC$ and $NC$ data 
is presented. 
Assuming that the shape of the SNO CC energy spectrum is undistorted
 and using the information provided by SNO we obtain, for the fraction of 
 electron neutrinos remaining in the solar beam at energies $\gsim 5$ MeV:
$$\phi_{CC}/\phi_{NC}=0.34^{+0.05}_{-0.04},$$
which is nominally $\sim 13\sigma$ away from the standard value ($\sim 1.8\pm 0.1$).
 The fraction  of oscillating neutrinos
which into active ones is computed to be:
$$
(\Phi_{NC}-\Phi_{CC})/(\Phi_{SSM}-\Phi_{CC})=0.92^{+0.39}_{-0.20}
$$
nearly $5\sigma$ deviations from the pure sterile oscillation case. The data is 
still compatible with an important fraction of sterile component in the 
solar beam (up to 20\% of the total).

In the framework of two active neutrino oscillations we determine individual 
neutrino mixing parameters and their errors in the region of no spectrum 
distortion ($\Delta\langle T_e\rangle <1\%$), 
we obtain 
$$
 \Delta m^2= 4.5^{+2.7}_{-1.4}\times 10^{-5} \eV^2,\quad \tan^2\theta=  0.40^{+0.10}_{-0.08}.
$$ 
This is in agreement with the best $\chi^2$ solution in the LMA region.

\vspace{0.3cm}
PACS:
\vspace{0.7cm}

{\scriptsize \noindent 
$\star$ email: paul@lcm.mi.infn.it, vito.antonelli@mi.infn.it,
r.ferrari@mi.infn.it,
 marco.picariello@mi.infn.it, torrente@cern.ch}

\end{abstract}

\newpage


{\bf 1.} Observation of neutral current neutrino interactions on deuterium in 
 the SNO experiment has been recently presented \cite{sno2002a,sno2002b}.
Using the NC, ES and CC reactions and assuming the ${}^8 B$ neutrino 
 shape predicted by the SSM, the electron and active non-electron component
 of the solar flux at high energies ($\gsim $5 MeV) are obtained.
 The non-electron component is found to be $\sim 5\sigma $ greater than 
zero, the standard prediction, thus providing the strongest evidence so 
 far for flavour oscillation in the neutral lepton sector:
the agreement  of the total flux, provided by the NC measurement
 with the expectations 
 implies as a by-product the confirmation of the validity of the
 SSM~\cite{turck,bpb2001,bp95}.

The $SNO$ experiment measures the \boron Solar neutrinos
 via the reactions~\cite{SNO,Boger:2000bb,Barger:2001pf,Bahcall:2001hv,Bahcall:2001hv}:
1) Charged Current (CC): $\nu_e + d\rightarrow 2 p+e^-$,
2) Elastic Scattering (ES): $\nu_x + e^-\rightarrow \nu_x+e^-$.
3) Neutral Current(NC): $\nu_x + d\rightarrow p+n+\nu_x$.
The first reaction is sensitive exclusively to electron neutrinos.
The second, the same as the one used in SuperKamiokande (SK),
 is instead sensitive, with different efficiencies, to all flavours.
Finally the NC reaction is equally sensitive to all active neutrino 
species.
 
The  results presented recently by $SNO$ on Solar neutrinos 
\cite{sno2001} 
 confirm and are consistent with previous evidence from SK and other
experiments~\cite{homestake,Fukuda:1999rq,Fukuda:1998fd,Fukuda:1998fd,sage1999}.
The CC, ES and NC global and day and night 
fluxes  presented in Refs.~\cite{sno2002a,sno2002b} 
, summarised in Table~\ref{tableratios},
are derived under 
 the assumption that the ${}^8 B$ spectral shape is not distorted from 
the SSM prediction. 
With this assumption the SNO collaboration  checks 
the hypothesis of non-oscillation, or zero $\phi_{\mu+\tau}$ flux.

It would be advantageous to use this assumption of non-distortion 
for several reasons. Not       
only in general terms of simplicity and logical economy
but also because with it a much higher statistical 
accuracy and power of prediction can be obtained (compare the total NC 
fluxes obtained with and without the distortion hypothesis obtained in 
\cite{sno2002a}).
Using  these fluxes is  appropriate for the calculation of constraints 
on mixing parameters in theoretical models where such spectrum distortion is 
negligible, this is true in particular for ample regions of oscillation 
 space favoured by all previous data (with or  without previous SNO CC data). 

Usually,  the best fit to the data has been routinely obtained 
in the LMA region
(see  \cite{torrente2001} and references therein), 
the qualitative explanation for that being the fact 
that just in that region the observed rather undistorted SK spectrum can 
be optimally adjusted.
The spectrum distortion of the oscillating solutions for SNO in the LMA region 
has been explicitly found to be negligible \cite{Bahcall:1996bw}.
Quantitavely,
the main information content of the shape of the observable spectrum is
summarised by the first moment of the distribution, the average 
spectrum energy. 
 For the SSM case and in absence of oscillations, it is 
found in Ref.\cite{Bahcall:1996bw} that the average kinetic energy 
$\langle T_e\rangle=7.658\pm 0.006$ MeV. 
This has to be compared 
with the  expected value  for a typical, 
non distorting, LMA oscillating 
solution $\langle T_e\rangle=7.654$ MeV
and with the far values of the distorted 
 SMA (7.875 MeV) and VAC (8.361 MeV) solutions.

In this work we present an up-to-date analysis of all available Solar neutrino 
evidence including latest SNO results in the most simple framework.
First we will reobtain some model independent results which put in a 
quantitative basis  the extent of the deviations with respect to the 
 standard non-oscillating case and the relative importance of active/sterile 
 oscillations.
Second we will obtain allowed areas in parameter space in the framework of 
active two neutrino oscillations from a standard statistical analysis. Individual 
values for $\Delta m^2$ and $\tan^2\theta$ with error estimation will be 
obtained from the analysis of marginal likelihoods. 
In this statistical analysis we include  all available data from SK, Homestake 
and Gallium experiments. From SNO we include the latest results on global 
 day and night fluxes, we make use in particular of: a) the total ${}^8\rm B$ flux 
as measured by the NC reaction and b) the electron neutrino day-night global 
asymmetry. 
The main conclusion of our analysis to be presented below is that it is already possible to 
determine at present active two neutrino oscillation parameters with relatively 
good accuracy.

\vspace{0.4cm}
{\bf 2.} Different quantities can be defined in order to make  the evidence for 
disappearance and  appearance of the neutrino flavours explicit. 
Letting alone the SNO data, from the three fluxes measured by SNO is possible to 
define two useful ratios, deviations of these ratios with respect to  their 
standard value are powerful tests for occurrence of new physics.
Here we compute the
values for $\phi_{CC}/\phi_{ES}$ and 
$\phi_{CC}/\phi_{NC}$  being specially careful with the treatment of the 
correlations on the uncertitudes, the inclusion or not of these correlations
can affect significantly their results (see table II in Ref.\cite{sno2002a}).
For the first ratio, from the value from SNO rates\cite{sno2002a} we obtain
$$ \frac{\phi_{CC}}{\phi_{ES}}=0.73^{+0.10}_{-0.07}, $$
 a value which is  $\sim$ 2.7 $\sigma$ away from the  no-oscillation expectation value of one.
The ratio of CC and NC fluxes gives the fraction of electron neutrinos remaining 
in the solar neutrino beam, our value is:
$$\frac{\phi_{CC}}{\phi_{NC}}=0.34^{+0.05}_{-0.04},$$
this value is nominally many standard deviations ($\sim 13 \sigma$) away  from the 
 standard model case  ($\phi_{CC}/\phi_{NC}=1.88 \pm 0.08$ 
from Ref.\cite{Bahcall:1996bw}).

Finally, if in addition to SNO data  we consider the 
 flux predicted by the solar standard mode one can define, 
following   Ref.\cite{Barger:2001zs},  the 
quantity $\sin^2\alpha$, the fraction of 
''oscillation neutrinos which oscillated into active ones'', again using 
 the SNO data and fully applying systematic correlations 
(see table 2 in Ref.~\cite{sno2002a}), we find the 
following result:
$$
\sin^2\alpha=\frac{\Phi_{NC}-\Phi_{CC}}{\Phi_{SSM}-\Phi_{CC}}=0.92^{+0.39}_{-0.20}.
$$
The SSM flux is taken as the ${}^8\rm B$ flux predicted in Ref.\cite{bpb2001}.
Note that, although consistent with it, this result differs significantly from the number obtained in 
Ref.\cite{Barger:2001zs}, this is due to the introduction of 
 systematic correlations in our calculation. 
The central value is clearly below one (only-active oscillations).
Although electron neutrinos are still allowed to oscillate into 
sterile neutrinos the hypothesis of transitions to {\em only} sterile 
 neutrinos is rejected at nearly $5\sigma$, this significance  would be reduced  if we consider applying
 a 1-sided analysis to avoid non-physical values.


\vspace{0.4cm}
{\bf 3.} Our determination of neutrino oscillations in Solar
 and Earth matter and of the expected signal in each experiment  follows
 the standard methods found in the literature~\cite{torrente}. 
As is explained in detail in Ref.\cite{torrente2001},
for this analysis
we completely solve numerically the neutrino equations of evolution for all 
 the oscillation parameter space.
The survival probabilities for an electron neutrino, produced in the 
 Sun, to arrive at the Earth  are calculated in three steps.
The propagation from the production point to sun's surface is computed 
 numerically in all the parameter range 
 using the electron number density $n_e$ given by the 
 BPB2001 model~\cite{bpb2001} averaging over the production point.
The propagation in vacuum from the Sun surface to the Earth is computed 
 analytically. The averaging over the annual  variation of the orbit
 is also exactly performed using simple Bessel functions.
To take the Earth matter effects into account, we adopt a
 spherical model of the Earth  density and chemical composition.
In this model, the Earth is divided in eleven  radial density
 zones~\cite{earthprofile}, in each of which  a polynomial
 interpolation is used to obtain the electron density.
The gluing of the neutrino propagation in the three different 
 regions is performed exactly using an evolution operator
 formalism~\cite{torrente}.
The final survival probabilities are obtained from the corresponding
 (non-pure) density matrices built from the evolution operators in 
 each of these three regions.
The  night quantities are obtained using appropriate weights 
 which depend on the neutrino  impact parameter and the 
 sagitta distance from neutrino trajectory  to the Earth center,
 for each detector's geographical location.

The expected signal in each detector is obtained by convoluting neutrino 
 fluxes, oscillation probabilities, neutrino cross sections and detector 
 energy response functions. We have used 
 neutrino-electron elastic cross sections which include radiative
 corrections~\cite{sirlin1994}.
Neutrino cross sections on deuterium needed for the computation of the 
 SNO measurements are taken from~\cite{nakamura2001}.
Detector effects are summarised by the respective response functions,
 obtained by taking into account both the energy resolution and the detector
 efficiency.
The resolution function for SNO is that given in~\cite{sno2002a,sno2002b,snohowto}.
We obtained the energy resolution function for SK using the data
 presented in~\cite{Nakahata:1999pz,skthesis,sakurai}. 
The effective threshold efficiencies, which take into account the live 
 time for each experimental period, are incorporated into our
 simulation program. They are obtained from~\cite{Fukuda:2001nj}.

The statistical significance of the neutrino oscillation hypothesis
 is tested with a standard $\chi^2$ method which is explained in 
 detail in Ref.\cite{torrente2001}.
Our present analysis is based on the consideration of the 
following $\chi^2$ quantity made of three well differentiated pieces:
\begin{eqnarray}
  \chi^2_{\rm }&=& \chi^2_{\rm glob}+\chi^2_{\rm spec-sk}+\chi^2_{\rm SNO }.
\end{eqnarray}
The contribution of SNO to the $\chi^2$ is given by
\begin{eqnarray}
  \chi^2_{\rm SNO }&=&
\left( \frac{\alpha - \alpha^{\rm th}}{\sigma_\alpha}\right)^2+
\left( \frac{A_e^{th} -A_e^{exp} }{\sigma_A}\right)^2,
  \label{chiall}
\end{eqnarray}
We have introduced the flux normalisation factor $\alpha$ 
with respect to the SNO NC flux whose central and error values are 
given in Table~\ref{tableratios}. This flux normalisation will be used 
below as a scale factor for the SK spectrum.
The 
quantity $A_e$ is the asymmetry on the day and night electron 
neutrino rates
extracted from the SNO CC, ES and NC data, imposing the condition 
$A_{tot}=A_{e+\mu\tau}\equiv 0$, as predicted by active-only models \cite{sno2002b}.
The strong anticorrelation of $A_e$ and $A_{tot}$ makes the 
imposition of this constraint useful, thus reducing greatly the uncertainties 
with respect to the raw $A_{CC}$.

The definition of the $\chi^2_{\rm glob}$ function is the following:
\begin{eqnarray}
  \chi^2_{\rm glob}&=& ({\vR^{\rm th}-\vR^{\rm exp}})^T 
\left (\sigma^{2}\right )^{-1} ({\vR^{\rm th}-\vR^{\rm exp}})
\label{chi1}
\end{eqnarray}
 where $\sigma^2$ is the full covariance matrix made up of two terms, 
 $\sigma^2=\sigma^2_{\rm unc}+\sigma^2_{\rm cor}$. 
The diagonal matrix $\sigma^2_{\rm unc}$ contains the theoretical,
 statistical and uncorrelated errors while $\sigma^2_{\rm cor}$ contains 
 the correlated systematic uncertainties.
The ${\vR}^{\rm th,exp}$ are length 2 vectors containing
 the theoretical and experimental data normalised to the SSM 
 expectations.
The $2\times 2$ correlation matrix have been computed using standard 
techniques~\cite{lisimatrix,Goswami:2000wb}.
We have used data on the total event rates measured at  Homestake 
 experiment, at the gallium experiments SAGE~\cite{sage,sage1999},
 GNO~\cite{gno2000} and GALLEX~\cite{gallex} 
 (see Table~(\ref{tableratios}) for an explicit list of results and 
 references).
For the purposes of this work it is enough to summarise all the 
 gallium experiments in one single quantity by taking the weighted
 average of their rates.

For  the analysis of the SK energy spectrum,
 following closely the procedure assumed by the SK 
collaboration~\cite{Fukuda:2001nk,Fukuda:2001nj}, 
we consider  a $\chi^2$ function:
\begin{eqnarray}\label{chi2}
  \chi^2_{\rm spec}&=&\sum_{d,n} ({\alpha \vR^{\rm th}-\vR^{\rm exp}})^t 
\left (\sigma^{2}_{\rm unc}+\delta_{\rm cor} \sigma^{2}_{\rm cor}\right )^{-1}
 ({\alpha \vR^{\rm th}-\vR^{\rm exp}})+\chi^2_\delta\, .
\end{eqnarray}
Where the vectors of data and expectations \vR are defined as before.
We have introduced the SNO NC flux normalisation factor $\alpha$
given above and the correlation 
 parameter $\delta_{\rm cor}$
is assumed to be constrained  by the   last term in the sum:
$\chi^2_\delta= (\delta_{\rm cor} - \delta_{\rm cor}^{\rm th})^2/\sigma_\delta^2$.
The complete variance matrix is not a constant quantity.
It is obtained from combining the statistical variances with systematic
 uncertainties and dependent on this correlation parameter.
For each day and night spectrum the corresponding 19$\times$19 block
 correlation matrices are conservatively constructed assuming full
 correlation among energy
 bins
The components of the variance matrix are given by standard expressions
\cite{torrente2001} in terms of
statistic errors, bin-correlated and uncorrelated uncertainties.
The data and errors for individual energy 
 bins for SK spectrum has been obtained from Ref.~\cite{Fukuda:2001nj}. 
Other information from the SK results, such as the global day-night asymmetries,
is to a large extent already contained in the above-mentioned quantities. It is therefore
not included in our analysis and does not change the results presented further on.

\vspace{0.4cm}
{\bf 4.} To test a particular oscillation hypothesis against the parameters of the best fit 
and obtain allowed regions in parameter space we perform a 
 minimisation of the four dimensional function
 $\chi^2(\Delta m^2,\tan^2\theta,\alpha,\delta_{\rm cor})$. 
For $\delta_{\rm cor}=\delta_{\rm cor}^{\rm min},\alpha=\alpha_{\rm min}$, 
 a given point in the oscillation parameter space is allowed if 
 the globally subtracted quantity fulfils the condition 
 $\Delta \chi^2=\chi^2 (\Delta m^2, \theta)-\chi_{\rm min}^2<\chi^2_n(CL)$.
Where $\chi^2_{n=4}(90\%,95\%,...)=7.78,9.4,...$ are the quantiles for
 four degrees of freedom.
The $\chi^2$ summation now contains 41 bins in total: 3 from
 the global rates  and SNO asymmetry
 and 2$\times$19 bins for the  SK  day and night spectrums.

\begin{figure}[p]
\centering
\begin{tabular}{ll}
\psfig{file=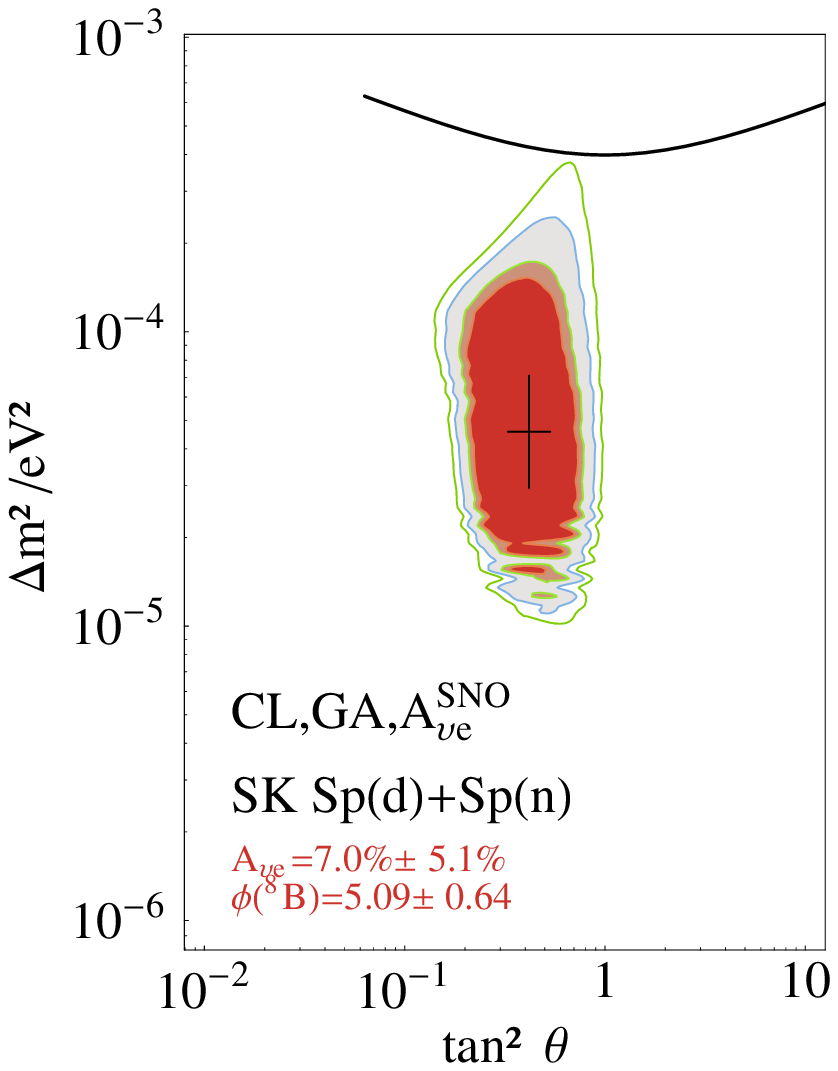,width=8cm} &
\psfig{file=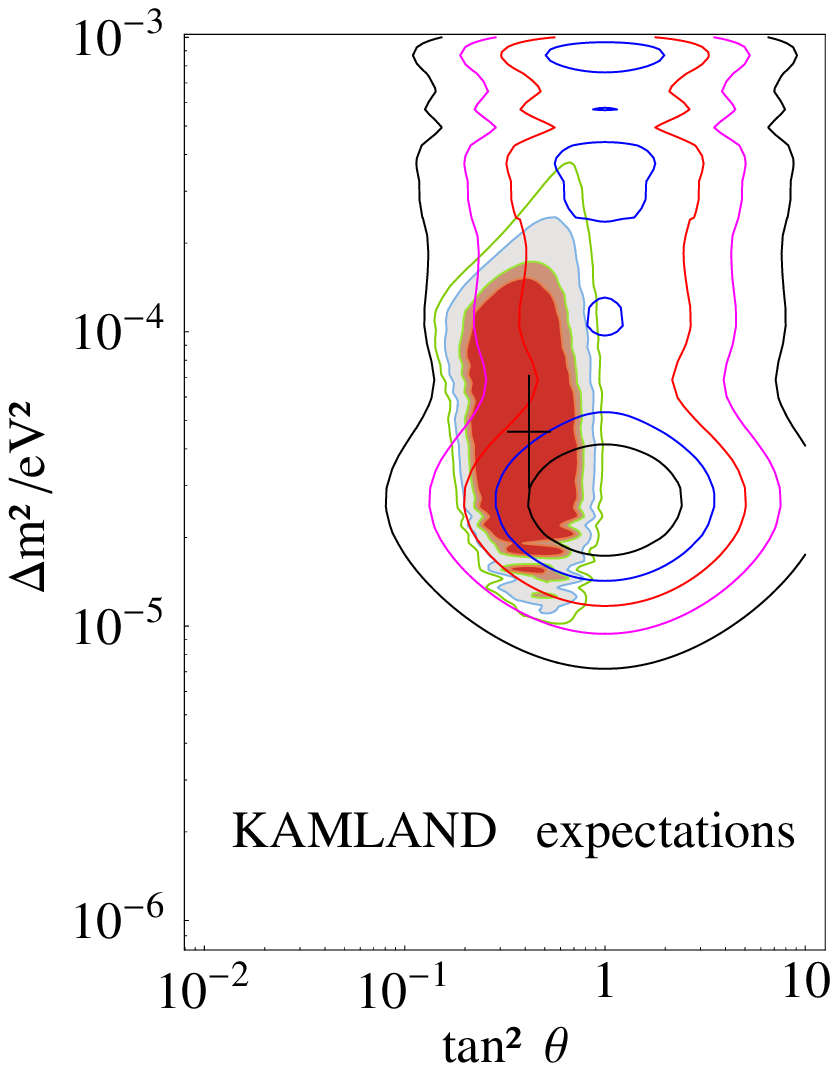,width=8cm}
\end{tabular}
\caption{\small
(Left)
Allowed areas in the two neutrino parameter space.
The point with error bars corresponds to best 
the results from fit to marginal likelihoods.
The coloured areas are the allowed regions at 
90, 95, 99 and 99.7\% CL relative to the absolute minimum.
The region above the upper thick line is excluded by the 
reactor experiments \protect\cite{chooznew}.
(Right) Superimposed to the  figure on the right, KamLAND constant 
signal contours normalized to the non-oscillation 
expectation. Contours, from inside to outwards, 
respectively at 0.4,0.5,0.6,0.7 and 0.8.
}
\label{f1}
\end{figure}

The results are shown in Fig.\ref{f1}(left) where we have generated acceptance 
contours in $\Delta m^2$ and $\tan^2\theta$.
In  Table~(\ref{table2})
 we present the best fit parameters or local minima 
 obtained from the minimisation of the $\chi^2$ function given in 
 Eq.~(\ref{chi1}).
Also shown are the values of $\chi^2_{\rm min}$ per degree of freedom
 ($\chi^2/n$) and the goodness of fit (g.o.f.) or significance level of each 
 point (definition of SL as in Ref.~\cite{pdg}).
 In Fig.\ref{f1}(right) we superimpose the global signal expected in the KamLAND
experiment from reactor electron antineutrinos. We observe that at the most 
favoured regions obtained before, the KamLAND expected signal is situated at 
 a intermediate region where high sensibility to both mass squared and mixing angle 
parameters is found. The experiment expects around $50\%$ 
of the non-oscillation signal, in this region it would suffice to reach a total 
error of $5-10\%$ (a quantity reachable after one year of 
 data taking) in the measurement to be able to confirm the SNO results.

\begin{figure}[p]
\centering
\begin{tabular}{ll}
\psfig{file=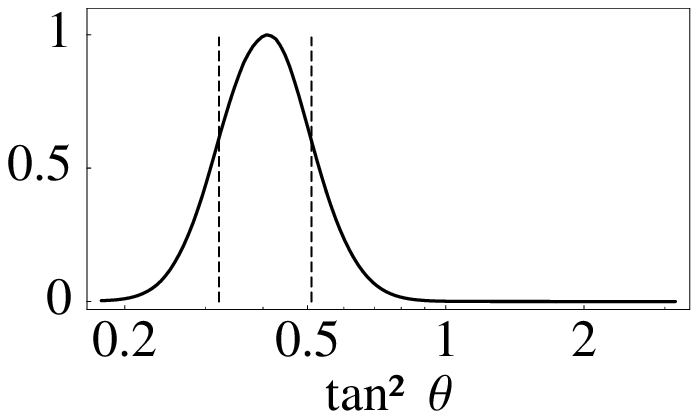,width=8cm,height=6cm} &\hspace{-1.5cm}
\psfig{file=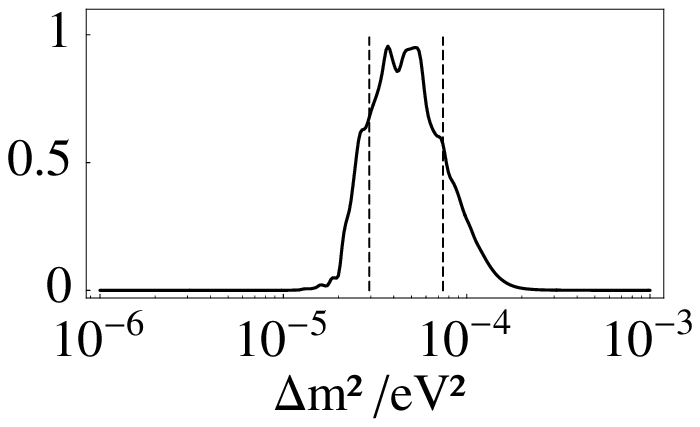,width=8cm,height=6cm} 
\end{tabular}
\caption{\small
Marginalized likelihood distributions for each of the
 oscillation parameters $\Delta m^2$ (right), $\tan^2 \theta$ (left)
 app
The specific signature of electron antineutrinos in proton
 containing materials is the inverse beta decay process:
 $\overline{\nu}_e+p\to n+e^{+}$, which produces 
almost isotropical   monoenergetic positrons with a relatively high 
cross section. Antineutrino events would contribute  to the 
 background to forward-peaked neutrino solar events.
earing in the $\chi^2$ fit.
The curves are in arbitrary units with normalization to the 
 maximum height.
Dashed lines delimit $\pm 1\sigma$ error regions around the maximum.}
\label{f2}
\end{figure}

In order to obtain concrete values for the individual oscillation parameters and 
estimates for their uncertainties, it is preferable
to study the marginalized parameter constraints.
It is justified to convert $\chi^2$ into likelihood
 using the expression ${\cal L}=e^{-\chi^2/2}$, this normalised marginal 
likelihood is plotted
 in Figs.~(\ref{f2}) for each of the oscillation parameters $\Delta m^2$ and $\tan^2\theta$.
For $\tan^2\theta$ we observe that the likelihood function is concentrated in a 
 region $0.2<\tan^2\theta<1$ with a clear maximum at $\tan^2\theta\sim 0.4$.
The situation for $\Delta m^2$ is similar.
Values for the parameters are extracted by fitting  one- or two-sided 
Gaussian distributions to any 
of the peaks. In the case of the angle distribution the goodness of fit of the 
Gaussian fit is excellent (g.o.f $>99.9\%$) even at far tail distances thus 
justifying the consistency of the procedure. The goodness of Gaussian fit to
the distribution in squared mass, although somewhat smaller, is still  good.
The values for the parameters appear in Table~\ref{table2}. They are 
fully consistent and very similar to the values obtained from simple 
$\chi^2$ minimisation.


In summary,
in this work we have present an up-to-date analysis of all available Solar neutrino 
evidence including latest SNO results in the most simple framework.
The direct measurement via $NC$ reaction on deuterium of $\phi_{^{8}{\rm B}}$ combined with the
$CC$ results has confirmed the neutrino oscillation hypothesis to $30\sigma$ according to 
our estimate. 
We have  obtained the allowed area in parameter space 
and individual 
values for $\Delta m^2$ and $\tan^2\theta$ with error estimation
 from the analysis of marginal likelihoods. 
We have shown that it is already possible to 
determine at present active two neutrino oscillation parameters with relatively 
good accuracy.
The KamLAND and specially the near future long baseline experiments will have 
a clear chance of first, confirming present  mixing  parameters obtained from 
 solar originated neutrinos and second, measuring first and second generation 
 mass and mixing parameters under laboratory-controlled conditions.

{\bf Acknowledgements.}
We  acknowledge the  financial  support of 
 the Italian MIUR, the  Spanish CYCIT  funding agencies 
 and the CERN Theoretical Division.
The numerical calculations have been performed in the computer farm of 
 the Milano University theoretical group.

\newpage

\begin{table}[p]
\centering
\scalebox{0.8}{
\begin{tabular}{|l|c|c|}
 \hline\vspace{0.1cm}
Experiment [Ref.] & $S_{SSM}$ & $S_{Data}/S_{SSM}\ (\pm 1\sigma)$
\\[0.1cm]\hline 
SK (1258d)     \protect\cite{Fukuda:2001nk}  & $2.32\pm 0.03\pm 0.08 $ & $ 0.451\pm 0.011$   \\[0.1cm]
Cl          \protect\cite{cl1999}  &  $2.56\pm 0.16\pm 0.16   $ & $0.332\pm 0.056$ \\[0.1cm]
SAGE        \protect\cite{sage1999,sage}  &  $67.2\pm 7.0 \pm 3.2  $ & $0.521\pm 0.067$ \\[0.1cm]
GNO-GALLEX  \protect\cite{gno2000,gallex}  &  $ 74.1\pm 6.7\pm 3.5 $ & $0.600\pm 0.067$  \\[0.1cm]
SNO data\protect\cite{sno2002a,sno2002b}:  & & \\[0.2cm]
\CCSNO     & $ 1.76\pm 0.06\pm 0.09$ & $0.348\pm 0.020$ \\[0.1cm]
\ESSNO     & $ 2.39\pm 0.24\pm 0.12$ & $0.473\pm 0.053$ \\[0.1cm]
\NCSNO     & $ 5.09\pm 0.44\pm 0.45$ & $1.008\pm 0.125$ \\[0.1cm]
$A_e (A_{TOT}\equiv 0)$  \protect\cite{sno2002b}  & $+0.070\pm 0.052$ & \\[0.1cm]
\hline
\end{tabular}
}
\caption{\small
Summary of data used in this work. 
The observed signal ($S_{SSM}$) and  ratios $S_{Data}/S_{SSM}$
 with respect to the BPB2001 model are reported.
The  SK  and SNO rates are in \fluxunit units.
The   Cl , SAGE and GNO-GALLEX measurements are in SNU units. 
In this work we use the combined results of SAGE and GNO-GALLEX: 
 $S_{\rm Ga}/S_{\rm SSM}$ ($\rm Ga\equiv SAGE$+GALLEX+GNO)=$0.579\pm 0.050 $.
The SSM \boron total flux is taken from the BPB2001 model~\protect\cite{bpb2001}:
 $\phi_{\nu}(\boron)=5.05 (1^{+0.20}_{-0.16}) \times 10^6$ cm$^{-2}$ s$^{-1}$
.}
\label{tableratios}
\end{table}

\begin{table}[p]
 \begin{center}
  \scalebox{0.95}{
    \begin{tabular}{lllll}
 Method  & & & & \\
\hline
 & & & & \\
A) Minimum LMA   & $\Delta m^2=5.44\times 10^{-5}\ \eV^2$ & $\tan^2\theta=0.40$    & $\chi_{m}^2$ = 30.8&  g.o.f.: 80\% \\
    & & & & \\
B) From Fit    & $\Delta m^2= 4.5^{+2.7}_{-1.4}\times 10^{-5} \eV^2$ & $\tan^2\theta=0.40^{+0.10}_{-0.08}$   &   &   \\
 & & & & \\
\hline
      \end{tabular}
}
 \caption{\small
Mixing parameters: 
A) Best fit in the LMA region and B) 
from fit to marginal likelihood distributions.}
  \label{table2}
 \end{center}
\end{table}

\end{document}